# Monochromatic high-harmonic generation by a Bessel-Gauss beam in periodically modulated media

O. Finke,[1,2,+] J. Vábek,[1,+] M. Dvořáček,[1,2] L. Jurkovičová,[1,2] M. Albrecht,[1,2] H. Ovcharenko,[1] O. Hort,[1] and J. Nejdl[1,2,*]

[1]*ELI Beamlines Facility, The Extreme Light Infrastructure ERIC, Za Radnicí 835, 252 41, Dolní Břežany, Czech Republic*
[2]*Czech Technical University in Prague, FNSPE, Břehová 7, 115 19 Prague 1, Czech Republic*
[+]*these authors contributed equally to this work*
[*]*jaroslav.nejdl@eli-beams.eu*

**Abstract:** High harmonic generation (HHG) has become a multipurpose source of coherent XUV radiation used in various applications. One of the notable aspects of HHG is its wide spectrum consisting of many harmonic orders. This might represent a bottleneck in HHG utility for applications requiring a single wavelength. We propose a method to generate radiation consisting of a single high-order harmonic frequency employing Bessel-Gauss driving beam and a periodically modulated gaseous medium. We validate it by analytical calculations and numerical simulations. Our method provides a way to generate monochromatic harmonic radiation directly from the source without the need for additional monochromatizing optics. Thus, it represents a substantial enhancement of the flux and simplification of the setup for numerous applications requiring monochromatic short-wavelength radiation.

## 1. INTRODUCTION

High-order harmonic generation (HHG) is a versatile tabletop source of coherent XUV radiation commonly using rare gases as a generation medium. Nonlinear interaction of the driver laser with the medium results in emission of radiation with a specific spectrum corresponding to multiples of a driver laser frequency which can range from vacuum ultraviolet (VUV) up to the X-ray part of the spectrum. The properties of this source provide wide utility in several research fields, notably in attosecond physics[1]; coherent diffractive imagining [2] or ptychography [3], and atomic, molecular, and optical sciences [4] or material sciences [5].

The spectrum of radiation from HHG consists of several harmonic orders, some applications of these sources, however, require only specific photon energy. In such cases, other harmonic orders can cause parasitic signals, and therefore need to be removed from the beam by, e.g., grating monochromators [6],[7]. These monochromators consisting of several reflective optical elements provide relatively low transmission, which decreases the flux of the desired harmonic typically below 30% [8],[9]. Recently, schemes for HHG using second or third harmonic of the fundamental laser to generate a spectrum with sparsely positioned harmonic orders have been reported [10][11][12][13]. The largely separated VUV harmonics can then be filtered using narrowband transmission filters or multilayered mirrors. Other approaches exploited wavefront control [14] or plasma resonances [15]. However, these approaches might not always be suitable, as the frequency upconversion of the laser driver lowers the harmonic cutoff, and narrowband optics reduce the versatility of the experimental setup. The aim of this study is, therefore, to design a HHG scheme providing highly monochromatic radiation (single harmonic frequency) for direct application without the need of further filtering or monochromatization.

Several advanced target designs for HHG have already used quasi phase-matching (QPM) to increase the yield of generated harmonics [16],[17],[18][19] by manipulating phase mismatch between the driver field and the generated one. Here we further expand this concept to suppress all harmonics except the optimized one. To achieve this, we employ Bessel-Gauss (BG) driving beams, which provide a strong geometrical phase shift along the focus. Furthermore, these beams form stable structures along the propagation axis [20]. We demonstrate that the strong geometrical phase, combined with a highly modulated longitudinal gas profile, provides suitable conditions for the generation of monochromatic radiation directly from the HHG source. Our design thus brings a novel mechanism to control the macroscopic HHG process by BG beams in the density-modulated media resulting in highly selective quasi-phase-matching.

The article is organized in the following way. First, the basic principle of monochromatic HHG is presented in Section 2. It is followed by the description of a detailed strategy to design an optimal generation scheme in Sec. 3; the stability of the scheme is analyzed in Section 4. The general presentation is accompanied by examples of monochromatizing harmonics within the range 20-30 in Ar by an 800 nm laser (as one of the typical conditions in HHG), and the 41st harmonic by 1030 nm in Ar (a case providing straightforward analysis as absorption is more uniform in the vicinity of this harmonic). Next, in Sec. 5, we use the scheme for the 401st harmonic generated by 1600 nm in He. This case is of particular interest as it falls within the water window, and absorption plays a minor role in the generating medium. These results are followed by a general discussion in Sec. 6.

## 2. BASIC PRINCIPLE OF MONOCHROMATIC HHG

The laser-atom interaction responsible for emission of XUV radiation [21] is happening throughout the medium, therefore it is usually beneficial, if the emerging $q^{th}$ order harmonic wave keeps a constant phase shift with the driver wave to add up constructively. In other words, the wave number mismatch between the driver and the $q^{th}$ order wave defined as $\Delta k_q = k_q - qk_1$ [22], where $k_q, k_1$ are corresponding wave numbers, should be reasonably small, or zero. In our model, we consider the generated harmonic beam as a plane wave, while the driving beam has nonzero wave vector contribution $k_{geo}$ due to its focusing, so that its on-axis wave number reads $k_1 = \frac{2\pi n_1}{\lambda_1} + k_{geo}$ with $\lambda_1$ denoting the laser wavelength and $n_1 = n_1(\lambda_1)$ the refractive index of the medium for the laser wavelength. Tailoring the geometrical phase of the driving beam and the refractive index of the medium is the core concept of our monochromatic HHG scheme.

We start with the refractive index in a periodically modulated gas target. For the initial simplification, let us assume that a medium with constant pressure alternates with empty spaces. We denote the length of a single period $z_p$ (containing the medium and the empty space). The basis of our concept is that the phase mismatch of desired harmonic order $q$ accumulated within one period of a generating medium that reads

$$\Delta\varphi_q(z_p) = \int_0^{z_p} \Delta k_q dz, \qquad (1)$$

equals $2\pi n$, where $n$ is a positive integer. This ensures that radiation of the $q^{th}$ harmonic from a period adds up constructively with the radiation from the previous period. Meanwhile, the absolute phase mismatch accumulated in the part of the period with a generating medium cannot exceed $\pi$ to ensure coherent buildup of the harmonic signal here. The rest of the phase-mismatch (up to $2\pi n$) is then accumulated in the empty space (see Fig. 1).

In the configuration, where the phase mismatch of the $q^{th}$ harmonic after one period equals $2\pi n$, the neighboring harmonic's mismatch will depart from $2\pi n$ as (1) is a function of the harmonic order. Note that the contribution of the phase mismatch in the empty space is caused by the geometrical phase of the driving laser. Therefore, one needs to employ a driving beam with a large geometrical contribution of the on-axis wave number. For Gaussian beams, which are typically used for driving HHG, the geometrical phase evolves only from $-\frac{\pi}{4}$ to $\frac{\pi}{4}$ across the focus region which is too little to achieve a large enough phase shift in single period of the medium, if we want to employ several periods within the laser focus. BG beams, on the other hand, provide a strong geometrical phase [20] described by the on-axis wave number as

$$k_{geo} = -\frac{2\pi n_1}{\lambda_1}\zeta = -\frac{2\pi n_1}{\lambda_1}(1-\cos\theta), \tag{2}$$

where $\zeta$ is a relative difference from a plane wave. Such beams can be generated by focusing a hollow laser beam as depicted in Fig. 2. The parameter $\zeta$ relates to the half-divergence angle $\theta$ of the BG beam by $\zeta = 1 - \cos\theta \approx \frac{\theta^2}{2}$. The on-axis wave number of BG beams is very high and almost constant across the focus. This allows us to design the experiment with the required geometrical phase shift of the desired harmonic order.

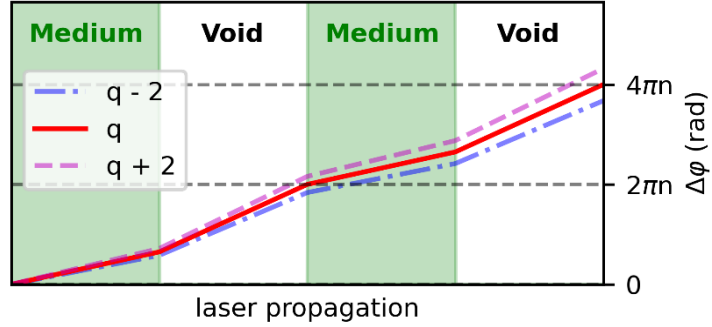

Fig. 1 – Example of evolution of the phase mismatch $\Delta\varphi_q$ in two medium periods. Green background highlights the presence of the gas. The phase mismatch for selected harmonic order $q$ evolves exactly with multiples of $2\pi n$ at the end of each period, while the neighboring orders get phase mismatch whose difference to $2\pi n$ increases with further propagation.

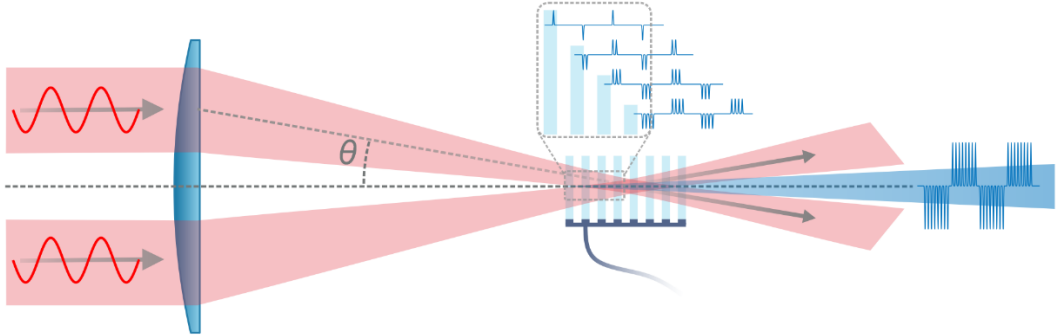

Fig. 2 – Schematic of the setup for monochromatic HHG. A hollow beam is focused into a periodically density-modulated gas target, where angle theta determines the geometrical phase according to (2). Blue schematic peaks denote temporal addition of attosecond pulses generated in consecutive medium segments (more details in text).

By employing multiple medium periods, the accumulated phase mismatch $\Delta\varphi_q$ of nondesired harmonic orders increases and generation of these orders is effectively suppressed. The rough estimate of the number of periods needed to suppress the neighboring harmonic orders reads $q/2n$ (see Appendix B for details). This value is accurate mainly in the low-absorption low-dispersion case, e.g., for HHG nonabsorbing gases, such as helium.

Even though we analyze mostly the HHG in targets with idealized "step-like" density modulation throughout this paper, as this can be calculated analytically (see Appendix A), the build-up of the harmonic signal can be also evaluated numerically for arbitrary density distribution of the medium as

$$E_q(L_{med}) = A_q \int_0^{L_{med}} \sigma(z) \, e^{i \int_0^z \Delta k_q(z')dz'} \, dz, \qquad (3)$$

where $A_q$ is the amplitude of the generated electric field of the $q^{th}$ harmonic, $\sigma(z)$ is density and $\Delta k_q(z)$ is the wave number mismatch. Note that the latter is a complex function of the spatial coordinate (accounting also for absorption of XUV in the medium).

In the first example, we will examine a case of the 29$^{th}$ harmonic (H29) generated by an 800 nm laser in argon. Figure 3 shows the signal of H29 and its neighboring harmonics as a function of propagation through the generating medium. For clarity and simplicity, we have chosen equal lengths of generating media and empty spaces. To show the robustness of our approach, we plot the results for the medium with a binary (steplike) density profile as well as a sinusoidal density profile. The phase-mismatch of the H29 after each period is equal to $2\pi$, i.e., $n = 1$, in this example of geometry configuration. Figure 3 clearly indicates that the signal of H29 is more than an order of magnitude stronger than neighboring harmonics H27 and H31.

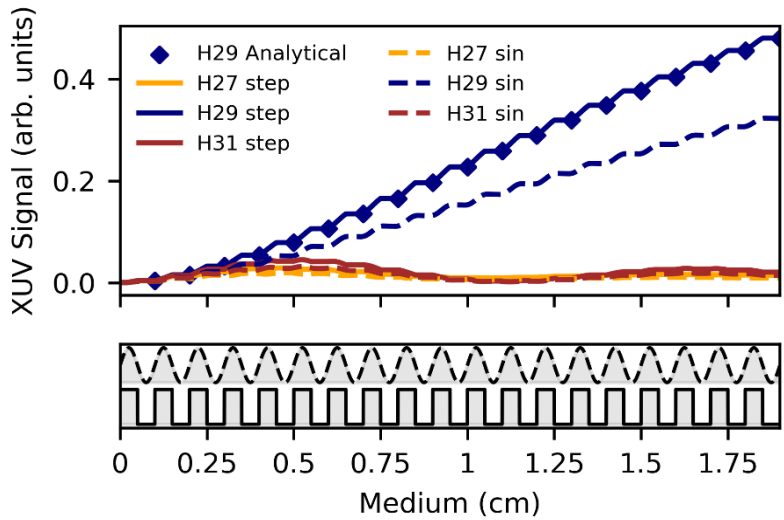

Fig. 3 – Evolution of the XUV signal generated in density-modulated argon using 800 nm driver as a function of the length of the medium. Signal of H27, H29, and H31 from binary (full line) and sinusoidal (dashed line) medium with peak argon pressure in the target of 100 mbar and divergence angle of the BG beam $\theta = 9.5$ mrad. Full and dashed lines are calculated using the numerical model (3) and blue diamonds correspond to analytical formula for binary medium (see Appendix A).

Considering our concept in the temporal domain, each of the medium segments is a source of an attosecond pulse train with characteristics typical for HHG with a single color driving field. A pulse train generated by the following segment is temporally advanced from the train generated by the previous one by $\frac{nT}{q} = \frac{n\lambda_q}{c}$ with $T$ denoting the period of the driving laser field and $c$ being the speed of light, because the driving field is effectively phase-shifted by $\frac{2\pi n}{q}$. The overall signal can, thus, be considered as an attosecond pulse train where the separation of peaks of the XUV electric field is an $n$ multiple of the $q^{th}$ harmonic period (as indicated in Fig. 2), which secures the significant spectral content of the $q^{th}$ harmonic and reduction of the other harmonic orders, if there are enough segments involved.

# 3. AN OPTIMIZED CONFIGURATION OF MONOCHROMATIC HHG

In the previous section, we presented the basic principle of monochromatic HHG by employing a Bessel-Gauss driving beam and density modulated periodic medium. In this section, we show a way to optimize the design of the monochromatic HHG scheme, i.e., we show how to determine the density profile of the generating medium and parameters of the driving BG beam. We will use a bottom-up strategy starting from an optimal design of generation in a single period of a binary medium using the analytical approach, i.e., assuming a steplike density profile for simplicity.

Every period of the generating target is divided into a segment with a medium of length $l_1$ and free space of length $l_2$ fulfilling $z_p = l_1 + l_2$. The goal is to maximize generation of the harmonic when the medium is present; therefore phase mismatch within the segment with the medium should be equal to 0 (phase matching is achieved there). Furthermore, the length $l_1$ and pressure $p$ should be chosen so that the total medium length-pressure product is not too high to enter the absorption limited generation regime. Therefore, the total effective length of the gaseous medium $L = Ml_1$, with $M$ denoting the number of periods, should be approximately

$$L \leq 3L_{abs}(p,q) \quad (4)$$

with $p$ denoting the pressure in the medium segment, where $L_{abs}$ is the absorption length of the $q^{th}$ harmonic order in the medium. Furthermore, the length of a single medium segment should always be

$$l_1 \ll L_{abs} \quad (5)$$

to achieve sufficient monochromaticity of the selected harmonic order. Depending on further conditions (harmonic order, absorption at this point), there is certain freedom to fulfill the optimizing conditions (4) and (5) to trade-off the contrast and the strength of the signal. By approaching the equality in condition (4), the signal strength of the selected harmonic increases to ~1/2 of the asymptotic absorption-limited maximum [23]; while prioritizing condition (5) enhances the monochromaticity of the generated radiation at the expense of harmonic signal strength. We found out that the number of periods in a medium corresponding to $\frac{q}{2n}$ provides a good rule of thumb for achieving reasonable signal monochromaticity, in particular for cases with low absorption (see Appendix B for details). In cases with significant absorption, i.e., where condition (4) approaches equality, there are generally, fewer medium periods needed.

Once $l_1$ and $p$ are set, the geometrical phase needs to be set up in a way to compensate for the effects of material dispersion and achieve perfect phase-matching in the segment with the medium. The optimum parameter (described above) of the BG beam thus reads (see Appendix A for further details of the derivation)

$$\zeta_{opt} = \frac{\sigma}{2}\left(\Delta\tilde{\alpha}_r - \eta\frac{e^2}{\epsilon_0 m_e \omega_1^2}\right), \quad (6)$$

where $\sigma$ is density, $\Delta\tilde{\alpha}_r$ is a real part of the difference of polarizabilities of neutrals between the driver and harmonic order, $\eta$ denotes the ionization degree of the gas, $e$ is electron charge, $\epsilon_0$ vacuum permittivity, $m_e$ electron mass, and $\omega_1$ angular frequency of the driving laser. In this design, the ionization is a parameter which is only restricted to the interval between 0 and the value corresponding to the complete plane-wave phase matching, $\eta_c$, of the studied gas. Because $\eta_c$ is typically a few percent, we do not correct $\Delta\tilde{\alpha}_r$ for the depletion of the neutrals (see the justification of this approximation in Appendix C). In this initial study, we assume that ionization is uniform in the region where the harmonics are efficiently generated. Generally, our optimization scheme works if $\eta < \eta_c$; we have chosen a fixed value of $\eta = \frac{\eta_c}{4}$ (see Appendix D for examples of other choices of $\eta$). The discussion on time-dependent ionization is in Sec. 6.

In the free-space segment of the target period, the geometrical phase of the beam is the only factor changing the phase-mismatch; therefore, the ratio of medium to vacuum segment lengths $\xi = l_2/l_1$ needs to be established as

$$\xi_{opt} = \frac{1}{\zeta_{opt}} \cdot \frac{2\pi n}{qk_0 l_1} \tag{7}$$

where $k_0$ is the vacuum wave number of the laser driver, to ensure the phase mismatch $2\pi n$ accumulated within the period. Note that the length of the medium $l_1$ generally differs from the length of free space $l_2$. By calculating $\zeta_{opt}$ and $\xi_{opt}$, the driving beam and the single period of the medium are fully defined.

At this point, the last parameter to determine is the number of periods. By increasing the constant $n$, fewer periods will be required to clearly pronounce only single harmonic order, on the other hand; it provides strain on the required target length.

The calculation of the signal of various harmonics after each period of a scheme optimized for monochromatic H41 in Ar driven by a 1030 nm laser shows that in the absorption-limited case ($L = 3L_{abs}$) there is a reasonable monochromaticity achieved already after 11 periods. The signal buildup is shown in Fig. 4. The dispersive properties are tabulated in [24] and [25], see [26] for all the details of the implementation.

We define the contrast of a given harmonic by the ratio of its signal and the signal of the strongest harmonic in its neighborhood. Therefore, various designs can be conveniently evaluated by comparing the total signal of the selected harmonic and its contrast.

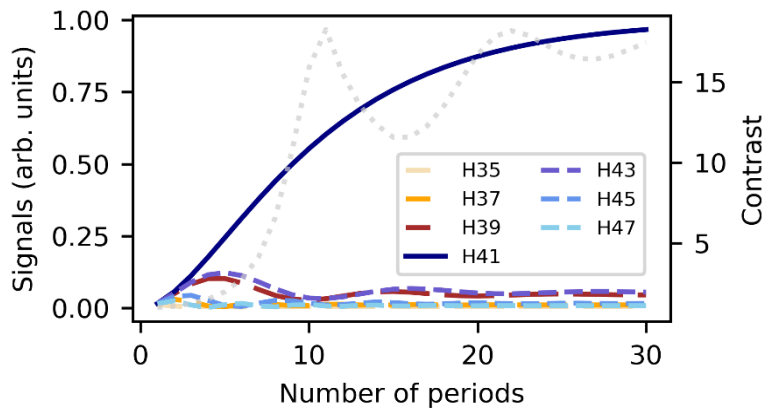

Fig 4 - Harmonic signal as functions of number of periods $M$ in a setup optimized for H41 generated in Ar by a 1030 nm laser. The design parameters are following: $l_1$=1 mm; $l_2$=0.87 mm; $p$=120 mbar; $\eta$=0.7%; and $\theta = 7.6$ mrad (n=1). Contrast of H41 is plotted with gray dashed line. (Note that the *arb. units* of the signal are related to the asymptotic value of perfectly phase-matched (absorption-limited) optimized harmonic generation. This definition of *arb. units* is applied in all the figures.)

Considering the design of monochromatic H41 generation with parameters shown in the caption of Fig. 4 as the reference case, we can elaborate upon other possible designs. The first alternative design has two-times shorter segments of the medium (while keeping all the other parameters fixed), which doubles the maximum contrast with respect to the reference case after 14 periods, while the total signal of H41 is 40% lower in that case [Fig. 5(a)]. Another interesting case is similar to the reference case, with doubled empty segments (again keeping all the other parameters fixed) to reach phase mismatch of H41 after each period to be $4\pi$, i.e., a case with *n*=2. A very high contrast of ~40 is achieved already after seven periods with lower H41 signal strength [Fig. 5(b)]. These examples represent a general trade-off between the contrast (monochromaticity) and the signal strength.

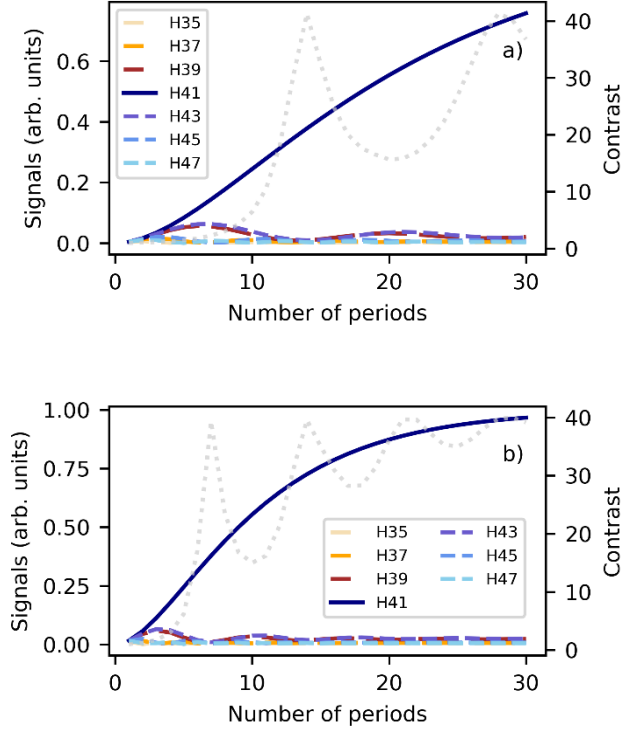

Fig. 5 - Variations of the design for monochromatic H41 in Ar driven by 1030 nm laser: a) two-times shorter segments b) n = 2. Setup parameters are following a) $l_1$=0.5 mm; $l_2$=0.87 mm; *p*=120 mbar; $\eta$=0.6%; and $\theta = 7.6$ mrad (n=1), and b) $l_1$=1 mm; $l_2$=1.74 mm; *p*=120 mbar; $\eta$=0.6%; and $\theta = 7.6$ mrad (n=2) b). Contrast of H41 is plotted with gray dashed line.

## 4. STABILITY OF THE METHOD

An important parameter of the monochromatic HHG is its stability with respect to various experimental parameters that might differ from the design values due to manufacturing or alignment errors. We have investigated the stability of the reference case of monochromatic generation of H41 designed in the previous section (using 11 periods of the medium) with respect to deviations of gas pressure in the target, geometrical phase of the driving beam (divergence angle of the BG beam), and ionization degree of the medium. The results are depicted in Fig. 6.

When the real experimental parameters depart from the design values, the easiest way to ensure fulfillment of the principal condition of monochromatic HHG, (i.e. the $2\pi n$ phase shift of the desired harmonic after one period) is usually to adjust the pressure in the target. Changing the pressure allows us to shift our scheme within the parametric space ($\theta$, pressure, ionization) into the region with a strong monochromatic signal. The differences between these regions for signals [Figs. 6(a) and 6(c)] and contrasts [Figs. 6(b) and 6(d)] also illustrate the trade-off between the contrast and the signal.

Consequently, one can conveniently optimize an experiment with the gas pressure to compensate for parameters that are difficult to tune experimentally, such as the driving beam divergence and ionization degree.

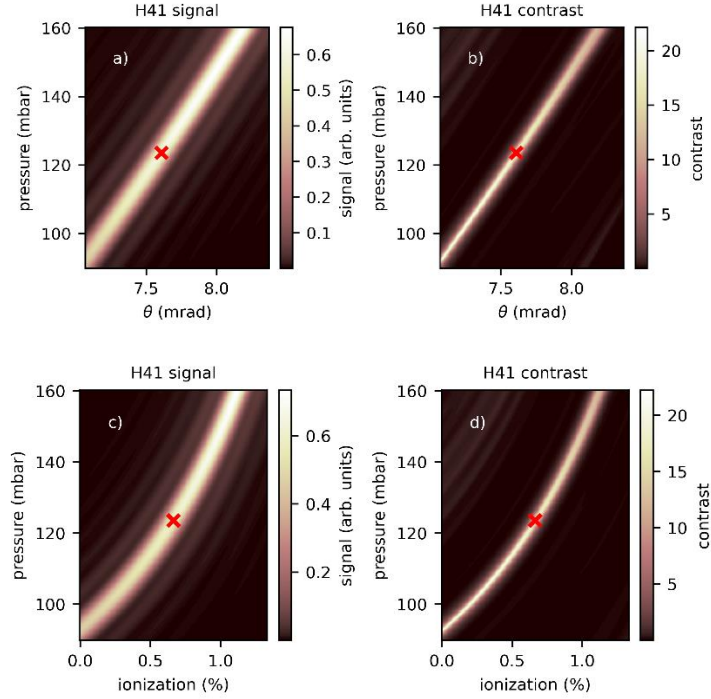

Fig. 6 - Signals (a,c) and contrasts (b,d) of the H41 as function of pressure in the medium, divergence angle of the BG beam $\theta$ (a,b), and degree of ionization (c,d). The central points of each figure indicated by the red cross correspond to the optimal designed values according to (6) and (7). Further examples of this case with ionization degree set to $\eta=0.5\eta_c$ and $\eta=0.75\eta_c$ can be found in Appendix D.

## 5. MONOCHROMATIC HHG IN THE WATER WINDOW

The monochromatic HHG setup can be designed for any type of medium. Generally, the selection mechanism works the best if the HHG is dominated by phase matching and not by absorption. As a next example of optimized monochromatic generation, we present the 401$^{st}$ harmonic of a 1600 nm laser ($\lambda_q \approx 4\ nm$) generated in helium, which falls into the water window wavelength range. Setting $l_1 = 200\ \mu m$ and $p = 100\ mbar$ at the beginning of the optimization procedure, the obtained harmonic signal (Fig. 7) is further away from the absorption limited optimum, because of the very low XUV photoionization cross section of helium (L~0.16 $L_{abs}$, assuming 200 periods). This is why this generation scheme can provide very high contrast of more than three orders of magnitude. The contrast of the optimized configuration is, indeed, mainly given by fulfillment of condition (5), as also seen in Eq. (B1), where the ratio $\frac{l_1}{L_{abs}}$ is the only parameter defining the contrast of the harmonic assuming a large enough number of periods.

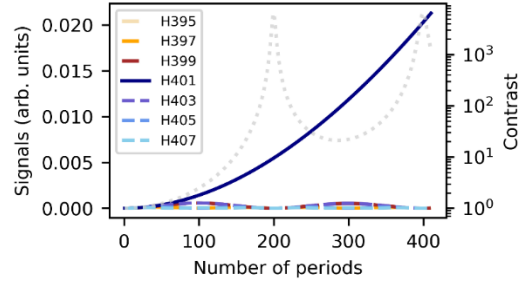

Fig. 7 - Signal of H401 of 1600nm driver laser and neighboring harmonic orders as a function of number of target periods *M* together with the contrast of H401. The parameters are following: $l_1$=0.2 mm; $l_2$=0.12 mm; $p$=100 mbar; $\eta$=0.34%; and $\theta = 7.61$ mrad (n=1). Contrast of H401 is plotted with gray dashed line (in logarithmic scale).

Due to the large number of periods needed for monochromatizing such high harmonic order it is worth considering a configuration with higher-order phase mismatch, i.e. with n=2 or n=3. The evolution of the harmonic signal along the periodic target for three configurations with n=1, 2, and 3 together with the contrast of the H401 is shown in Fig. 8.

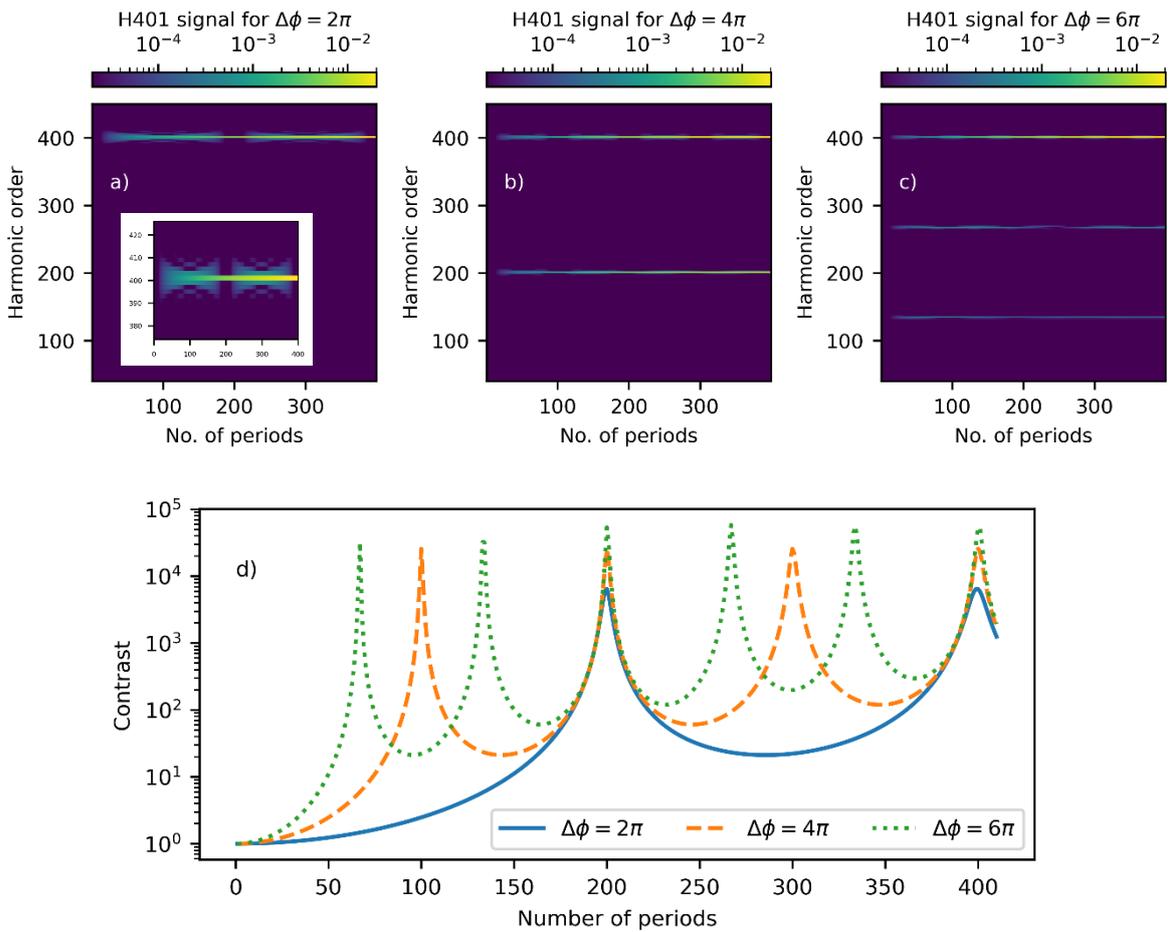

Fig. 8 - Signal of various harmonics as function of number of periods for experimental configuration described in caption of Fig. 7 (a) and for the same configuration but with $l_2$=0.24 mm (n=2) (b), and $l_2$= 0.36 mm (n=3) (c). Contrast of H401 as function of number of periods for the three phase mismatch configurations (d).

These results indicate that, in general, the harmonic orders close to integer multiples of q/n are enhanced, which can give rise to subharmonic frequencies of the desired harmonic in the case of n>1. This is understandable, as while the harmonic $q$ gains phase mismatch $2\pi n$ after one period, the harmonic $q/n$ reaches the phase mismatch $2\pi$, or, more generally, harmonic $mq/n$ gains phase mismatch $2\pi m$ where $m$ is an integer. In fact, this might reduce the strict monochromaticity for higher order phase-mismatches. In some cases, this monochromatization might be sufficient, if the subharmonic peaks are spectrally separated allowing for their removal by a single high-pass filter. Experimentally, this can be achieved, e.g., with a state-of-the-art multilayer mirror, as the enhanced harmonics are spectrally considerably separated (200 harmonic orders for n = 2 and > 130 orders for n = 3). Therefore, we believe the schemes with higher values of $n$ broaden our parameter palette as these configurations might be practical due to less stringent experimental requirements mainly a lower number of medium periods needed.

## 6. DISCUSSION

The examples presented in the manuscript demonstrate the ability to analytically find the optimal parameters in the binary medium given by (6) and (7). The same strategy holds for any periodic density profile of the medium that provides sufficient modulation to reach required phase differences (see Appendix A). For example, Fig. 9 shows the analogical design as presented in Fig. 4 by replacing the binary medium by 11 "Gaussian jets", defined by $p(z) = p_0 \exp\left(-\left(\frac{z-z_i}{a}\right)^2\right)$, where $z_i$ is the position of the jet and $a$ characterizes its width. The optimal values of the geometrical phase and jets' spacing are found numerically to optimize the signal of H41 (see corresponding Jupyter notebook in [26] for details). It shows that the optimization works similarly as in the idealized case, the contrast remains similar and the signal decreases by only 13 % with respect to the analytical case with a binary medium.

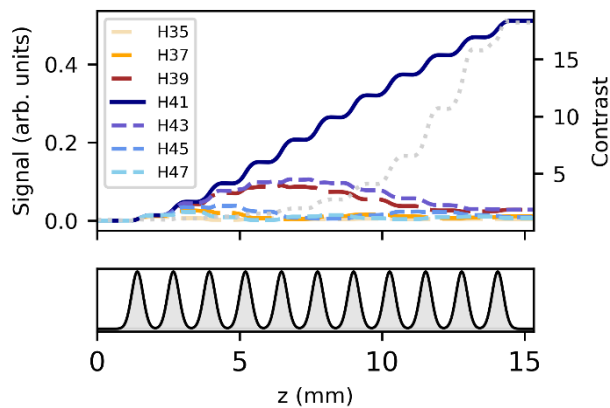

Fig. 9 - The signal of H41 and its contrast for the medium with "Gaussian jets" (the density profile is depicted below the plot) instead of binary profile of Fig. 4. The medium period is ~1.3 mm, $a \approx 0.3$ mm, and peak pressure $p_0 = 120$ mbar (the amount of gas in each period is the same as in Fig. 4.), η=0.7%. The optimized BG beam divergence is $\theta_{\text{opt.}} = 9.20$ mrad.

Although in practice it might be difficult to set all the experimental parameters exactly, a similar design principle as described in Sec. 3 can be applied experimentally. This way, one optimizes the XUV signal with a single segment. Segment number and spacing are then to be tuned for the desired contrast.

Ionization degree is, unlike the other experimental parameters, a dynamic variable that might change during the passage of the driving pulse and reduce the overall contrast of the required harmonic order.

Pressure tuning cannot be used efficiently in this case as it is a static dispersion compensation method. Therefore, for the cases of rapid ionization of the generating medium, we propose to employ laser driving fields with BG beams and pulses with spatiotemporal coupling (STC) [26] that has decreasing divergence angle $\theta$ within the laser pulse to compensate for the generally increasing ionization degree of the medium. To give an example, we plot the contrast as a function of ionization degree and divergence angle of the BG beam in Fig. 10. In the cases of H41 generated by a 1030 nm laser driver in Ar and H401 generated by a 1600 nm laser in He, driving pulses with STC seem to be necessary to sustain monochromatic HHG. In contrast to that, the generation of H23 by an 800 nm laser in Ar shows much lower sensitivity to ionization degree and a sufficient level of contrast can be achieved even with standard pulses without STC. In general, the sensitivity of ionization degree decreases with decreasing wavelength of the laser driver and decreasing harmonic order, but also with the medium density [mathematically, this stems from the optimizing condition (A7) in Appendix A].

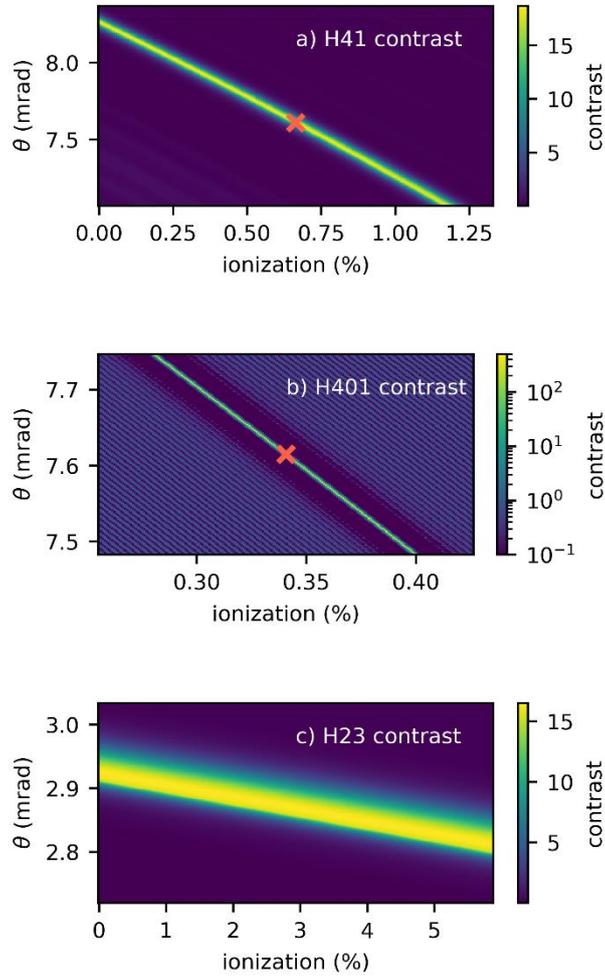

Fig. 10 - Contrast as function of ionization degree and divergence angle of the BG beam for H41 generated in Ar with 1030 nm laser (a), H401 generated in He with 1600 nm laser (in log scale) (b), and H23 generated in Ar by 800 nm laser (c). The red crosses indicate the design values of the schemes.

Moreover, it should be noted that monochromatic HHG with BG beams and pulses with STC would benefit from the possibility to adjust the group velocity in the focus [28], because the pulses in the BG beam without STC are superluminal as $v_g = \frac{c}{\cos\theta}$. The monochromatic HHG with BG beams and STC is, however, beyond the scope of this paper and it will be discussed in the following publication.

Experimental limitations of the monochromatic HHG lie mainly in fitting the desired number of medium periods in the region, where geometrical phase together with sufficient laser intensity is available, or in technological limitation of preparing the density-modulated medium with high enough modulation on a short length scale. However, we believe the example designs presented here are feasible with current manufacturing and laser technologies.

## 7. CONCLUSION

In this article, we introduced a novel method for a direct generation of purely monochromatic harmonic radiation using Bessel-Gauss beams and periodically density-modulated medium. We apply this method to monochromatize HHG with significant absorption (the generation in Ar) or negligible absorption (the generation in He), which shows the universality of the method. We believe that our generation scheme outperforms the standard HHG systems with monochromators as well as other monochromatizing methods, and provides strong monochromatic radiation required in many user applications. Furthermore, we provide our simulations in the form of a Jupyter notebook written in Python [26], for the readers to test our conclusions or to design a configuration suitable for their own application.

## A. APPENDIX – ANALYTICAL COMPUTATION OF THE SIGNAL

Here we present a simplified model of the harmonic generation by Bessel-Gauss beams (approximated by diffraction-free beams within our interaction region) in the discretized medium given by the rectangular profile of the pressure in the laser propagation direction $z$. This assumption permits us to find the analytic expressions of optimal parameters (6) and (7).

First, we find the phase mismatch using the diffraction-free beams and dispersion relations. Second, we derive the analytic expression of the harmonic signal. At the end, we will show that the main feature – the selection mechanism of a given harmonic – holds also for an arbitrary medium profile.

Let us introduce the geometry imprinted by the Bessel-Gauss beams. We consider the target composed from consecutive segments of the medium and vacuum of respective lengths $l_1$ and $l_2$. Within our assumption, the phase-mismatch along the propagation axis $z$ is retrieved from the diffraction-free beam profile $E_{IR}(z,\rho) = \exp(i\alpha z) J_0(\beta\rho)$, [29] where $\alpha$ and $\beta$ are linked with the dispersion relation by

$$\alpha^2 + \beta^2 = k^2, \tag{A1}$$

where $k$ is the wave number. We find how the geometrical phase $\alpha$ changes on the transition between the vacuum and the medium. Because of the continuity of the field on the interface, $\beta$ is equal on both sides and we have

$$\alpha_m = \sqrt{k_m^2 - k_v^2 + \alpha_v^2}, \tag{A2}$$

where the indices $v$ and $m$ denote vacuum and medium, respectively. We start our analytic considerations by the rectangular modulation of gas and vacuum segments, i.e., the gas density, $\sigma$, is constant in the gas segments and zero in the vacuum segments.

Now, we compute the harmonic signal from (3). First, we need to find the wave number mismatch

$$\Delta k_{x,q} = k_{x,q} - q\,\alpha_x, \qquad x \in \{m, v\}. \tag{A3}$$

Let us list the considered effects. In vacuum, the harmonics propagate as plane waves, $k_{v,q} = qk_0$, and the fundamental beam is affected only by the geometrical phase, $\alpha_v = k_0(1-\zeta)$ (see Eq. (2)). The situation is more complex in the medium. The harmonics are affected only by dispersion, $k_{m,q} = qk_0\sqrt{1+\sigma\tilde{\alpha}_q}$, where $\sigma$ is the density and $\tilde{\alpha}_q$ the polarizability for the $q$th harmonic. The fundamental beam is affected by the dispersion and free electrons; it gives $k_m = qk_0\sqrt{1+\sigma(\tilde{\alpha}_{IR} - \eta e^2/(\epsilon_0 m_e \omega_1^2))}$, the included quantities are described below (6). Considering the wave numbers, (A2), Taylorizing square roots, and neglecting $\zeta^2$-terms, the wave number mismatches are

$$\Delta k_{m,q} \approx qk_0\left(\frac{\sigma}{2}\left(\tilde{\alpha}_{IR} - \tilde{\alpha}_q - \frac{\eta e^2}{\epsilon_0 m_e \omega_1^2}\right) - \zeta\right), \tag{A4a}$$

$$\Delta k_{v,q} = -qk_0\zeta. \tag{A4b}$$

Using (A4a), we can find the signal at the end of the first gas segment

$$S_1 = \int_0^{l_1} A_q\, e^{i\Delta k_{m,q} z} dz = \frac{iA_q(e^{i\Delta k_{m,q} l_1} - 1)}{\Delta k_{m,q}}, \tag{A5}$$

with $A_q$ denoting the amplitude of the generated electric field of the $q^{th}$ harmonic. Because the dephasing is piecewise constant, its integration gives a continuous piecewise linear phase and the signal at the end of the $M^{th}$ gas segment is obtained as the signal from the first segment modulated by a geometrical sum:

$$S_M = S_1 \sum_{m=0}^{M} e^{im\Delta\varphi_q} = S_1 \frac{(e^{i\Delta\varphi_q M} - 1)}{(e^{i\Delta\varphi_q} - 1)}, \tag{A6}$$

$$\Delta\varphi_q = qk_0 l_1\left(\frac{\sigma}{2}\left(\tilde{\alpha}_{IR} - \tilde{\alpha}_q - \frac{\eta e^2}{\epsilon_0 m_e \omega_1^2}\right) - \zeta(1+\xi)\right), \tag{A7}$$

where $\xi = \frac{l_2}{l_1}$. Note that the stability and sensitivity of a scheme (discussed in Sections 4 and 6 and) are inferred from (A7). The selection condition is $\mathrm{Re}(\Delta\varphi_q) = 2\pi n$, where $n$ is an integer. Since this condition is achieved by the cancellations of different terms contained in (A7), the required relative precision imposed on the different quantities $(\sigma, \eta, \zeta, \xi)$ increases as they grow. In other words, the stability decreases if the aforementioned quantities increase.

Now, we find the analytic form of the XUV intensity:

$$|S_1|^2 = 4|A_q|^2 e^{-\frac{l_1}{2L_{abs}}} \frac{\sinh^2\left(\frac{l_1}{4L_{abs}}\right) + \sin^2\left(\frac{qk_0 l_1}{2}\left(\frac{\sigma}{2}(\Delta\tilde{\alpha}_r - \eta\mathcal{A}) - \zeta\right)\right)}{q^2 k_0^2\left(\frac{\sigma}{2}(\Delta\tilde{\alpha}_r - \eta\mathcal{A}) - \zeta\right)^2 + \frac{1}{4L_{abs}^2}}, \tag{A8}$$

$$|S_M|^2 = |S_1|^2 e^{-\frac{(M-1)l_1}{2L_{abs}}} \frac{\sinh^2\left(\frac{Ml_1}{4L_{abs}}\right) + \sin^2\left(\frac{Mqk_0 l_1}{2}\left(\frac{\sigma}{2}(\Delta\tilde{\alpha}_r - \eta\mathcal{A}) - \zeta(1+\xi)\right)\right)}{\sinh^2\left(\frac{l_1}{4L_{abs}}\right) + \sin^2\left(\frac{qk_0 l_1}{2}\left(\frac{\sigma}{2}(\Delta\tilde{\alpha}_r - \eta\mathcal{A}) - \zeta(1+\xi)\right)\right)}. \tag{A9}$$

where $L_{\text{abs}} = \frac{1}{2\text{Im}(k_{m,q})}$, and $\mathcal{A} = \frac{e^2}{\epsilon_0 m_e \omega_1^2}$. Note that the absorption comes only from the imaginary part of $\tilde{\alpha}_q$. Using the power-reduction for sin, the signal within the single gas segment, $|S_1|^2$, given by (A8) can be recast into equation (1) from [23], where $L_{\text{coh}} = \frac{\pi}{\left|qk_0\left(\frac{\sigma}{2}(\Delta\tilde{\alpha}_r - \eta_e\mathcal{A}) - \zeta\right)\right|}$. Equations (A8) and (A9) directly provide the optimizing conditions for a given harmonic $q$. The former equation gives (6) and the latter (7).

Finally, we show that this reasoning can be generalized for an arbitrary gas profile. Let us assume that the support of an elementary medium is confined within the distance $l_1$ and the gas segments are spaced by $l_2$. The signal at the end of the first gas segment is

$$S_1 = \int_0^{l_1} A_q(z)\, e^{\mathbf{i}\int_0^z \Delta k_m(z')dz'}\, dz. \tag{A10}$$

Next, we find the signal at the end of the second gas segment

$$\begin{aligned}
S_2 &= S_1 + \int_{l_1+l_2}^{2l_1+l_2} A_q(z) e^{\mathbf{i}\int_0^z \Delta k(z')dz'}\, dz = S_1 + \int_0^{l_1} A_q(z'+l_1+l_2) e^{\mathbf{i}\int_0^{z'+l_1+l_2} \Delta k(z'')dz''}\, dz'\\
&= S_1 + e^{\mathbf{i}\int_0^{l_1+l_2}\Delta k(z)dz}\int_0^{l_1} A_q(z) e^{\mathbf{i}\int_{l_1+l_2}^{z+l_1+l_2}\Delta k(z')dz'}\, dz\\
&= S_1 + e^{\mathbf{i}\int_0^{l_1+l_2}\Delta k(z)dz}\int_0^{l_1} A_q(z) e^{\mathbf{i}\int_0^z \Delta k(z''+l_1+l_2)dz''}\, dz\\
&= S_1 + e^{\mathbf{i}\int_0^{l_1+l_2}\Delta k(z)dz}\int_0^{l_1} A_q(z) e^{\mathbf{i}\int_0^z \Delta k(z')dz'}\, dz = S_1\left(1 + e^{\mathbf{i}\int_0^{l_1+l_2}\Delta k(z)dz}\right); (A11)
\end{aligned}$$

where we assume that both $A_q(z)$ and $\Delta k_m(z)$ depend on $z$ only through the density (this means $A_q(z) = A_q(\sigma(z))$ and $\Delta k_m(z) = \Delta k_m(\sigma(z))$) meaning that they are $(l_1 + l_2)$-periodic, which justifies the equalities within (A11). The same reasoning can be inductively repeated for adding more gas segments. It thus leads to the same geometric sum as (A6). The only difference is that the phase $\Delta\varphi_q$ is not expressed analytically and is replaced by

$$\Delta\varphi_q = \int_0^{l_1+l_2}\Delta k(z)\, dz = \int_0^{l_1}\Delta k_m(z)\, dz - qk_0 l_2 \zeta. \tag{A12}$$

The above derivation uses the assumption that the gas segments are separated by vacuum. However, the vacuum could be easily replaced by a residual nonzero gas density until the contribution to the XUV signal generated there, which is dephased from the effective gas segments, is small. In other words, the density modulation must be strong relative to the residual density.

In conclusion, the main monochromatizing mechanism stays unaffected for a general gas profile with a sufficiently strong density modulation. The difference is that the optimizing parameters cannot be generally expressed analytically as for the discretized medium.

## B. APPENDIX – NUMBER OF PERIODS AND CONTRAST

The contrast and its evolution with the number of gas segments is inferred from the analytic model (A9). We define the contrast of harmonics $q$ and $\tilde{q}$ after $M$ segments as $C_{q,\tilde{q},M} = \frac{\left|S_M^{(q)}\right|^2}{\left|S_M^{(\tilde{q})}\right|^2}$. Let us assume that the dispersion is the same for the harmonics $q$ and $\tilde{q}$. We start with the estimate of the number of periods

while the absorption is negligible ($L_{abs} = +\infty$). The theoretical contrast in this case is extremely high (see Fig. 7). The optimizing condition for the harmonic selection (7) leads to $\frac{k_0 l_1}{2}\left(\frac{\sigma}{2}\left(\tilde{\alpha}_{IR} - \text{Re}(\tilde{\alpha}_q) - \eta\mathcal{A}\right) - \zeta(1+\xi)\right) \to \frac{\pi n}{q}$. If we insert this value into the signal (A9) for the $\tilde{q}^{\text{th}}$ harmonic, the numerator gives $\sin^2\left(\frac{M\tilde{q}\pi n}{q}\right)$. The first zero of this function (i.e. the value for which the signal of the $\tilde{q}^{\text{th}}$ harmonic vanishes) is reached at $M = \frac{q}{|n|\Delta q}$, where $\Delta q = |\tilde{q} - q|$. Particularly, we find $M = \frac{q}{2}$ for neighboring harmonics and $n = 1$.

If the absorption plays a significant role (as in cases studied in Figs. 4 and 5), Eq. (A9) allows us to evaluate the contrast as a function of the absorption length. The contrast is

$$C_{q,\tilde{q},M} = \left(\frac{\sinh^2\left(\frac{l_1}{4L_{abs}}\right) + \sin^2\left(\frac{\tilde{q}}{q}n\pi\right)}{\sinh^2\left(\frac{l_1}{4L_{abs}}\right)}\right)\left(\frac{\sinh^2\left(\frac{Ml_1}{4L_{abs}}\right)}{\sinh^2\left(\frac{Ml_1}{4L_{abs}}\right) + \sin^2\left(M\frac{\tilde{q}}{q}n\pi\right)}\right) \xrightarrow{M\to+\infty} \frac{\sinh^2\left(\frac{l_1}{4L_{abs}}\right) + \sin^2\left(\frac{|n|\Delta q}{q}n\pi\right)}{\sinh^2\left(\frac{l_1}{4L_{abs}}\right)}. \quad (B1)$$

The limit $M \to +\infty$ provides the asymptotic contrast, which gives a good estimate of the expected contrast even for lower number of periods (see the difference in the scale of the contrast in Figs. 4 and 5). Since the optimal conditions are reached for all the arguments close to 0, the asymptotic contrast is $C_{q,\tilde{q},M} \approx 1 + 16\pi^2\left(\frac{n\Delta q L_{\text{abs}}}{q l_1}\right)^2 \approx 16\pi^2\left(\frac{n\Delta q L_{\text{abs}}}{q l_1}\right)^2$. It shows that the contrast increases with the ratio $L_{\text{abs}}/l_1$ and with the order of phase-mismatch $n$. As this model is fully analytic, these particular cases may be efficiently studied with simple tools (see the attached notebooks [26]).

## C. APPENDIX – PHASE-MATCHING MODEL: THE DEPOPULATION OF NEUTRALS AND IONIC CONTRIBUTION TO POLARIZABILITY

We have neglected the depopulation of neutrals due to ionization in our phase-matching model. In some models, such as in [30], the depopulation is taken into account, but the ions are not assumed to contribute to the polarizability. Here, we show that the difference is negligible for the small ionization degree assumed in our work. First, we show the difference between the two models. Then, we derive their error analytically while also considering the ionic contribution to the polarizability.

Figure 11 shows the signals and contrast calculated with our model and the one including the depopulation of neutrals. The latter case is modeled by replacement of $\left(\tilde{\alpha}_{IR} - \tilde{\alpha}_q\right)$ by $\left(\tilde{\alpha}_{IR} - \tilde{\alpha}_q\right)(1-\eta)$ in Eq. (A4a). Indeed, the differences are small.

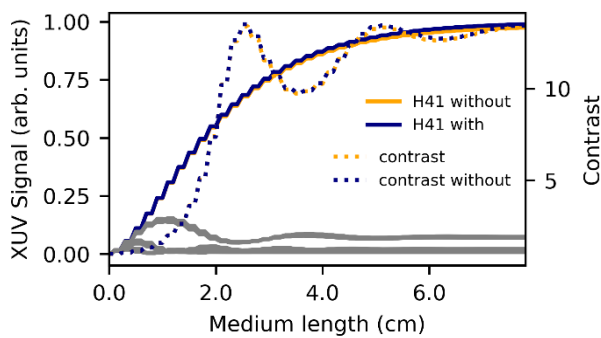

Fig. 11 - Comparison of signal and contrast evolution of H41 of 1030 nm driver with BG beam with divergence of $\theta = 7.4\ mrad$ in argon medium ($l_1 = l_2 = 1\ mm$, medium pressure 124 mbar, and $\eta = 0.6\%$). Blue curves represent the situation without included depopulation of neutrals, as in the phase-

matching model used throughout the manuscript, and green curves represent the model with included depopulation of neutrals.

Now we further analytically estimate the difference with and without assuming the depletion of neutrals. In both cases, whether neutrals are reduced by $(1 - \eta)$ or not, the models do not account for the dispersive properties of ions. Let us include them. Denoting $\Delta\alpha = \alpha_{IR} - \alpha_q$ the difference of polarizabilities of neutrals and $\Delta\alpha_+ = \alpha_{IR,+} - \alpha_{q,+}$ the difference of the polarizabilities of ions, we can write the wave number mismatch as

$$\Delta k_q = q k_0 \left[ \frac{\sigma}{2} (\Delta\alpha(1-\eta) + \Delta\alpha_+ \eta - \mathcal{A}\,\eta) - \zeta \right]. \quad (C1)$$

First, we can find the optimal value of ionization degree providing a perfect phase-matching for plane waves (without considering geometrical effects, i.e. $\zeta = 0$) as defined in [31]. In the complete model it reads

$$\eta_c = \frac{\Delta\alpha}{\mathcal{A}} \left( \frac{1}{1 + \frac{\Delta\alpha - \Delta\alpha_+}{\mathcal{A}}} \right) = \frac{\Delta\alpha}{\mathcal{A}} \left( 1 - \frac{\Delta\alpha - \Delta\alpha_+}{\mathcal{A}} + \left(\frac{\Delta\alpha - \Delta\alpha_+}{\mathcal{A}}\right)^2 - \left(\frac{\Delta\alpha - \Delta\alpha_+}{\mathcal{A}}\right)^3 + \cdots \right). \quad (C2)$$

The first correction, which gives the relative error, is of the order of $\frac{\Delta\alpha - \Delta\alpha_+}{\mathcal{A}}$, which is itself about a few percent because this value corresponds to the typical values of the optimal ionization for phase-matching.

Next, we use the same reasoning to express the contribution of the geometrical effect $\zeta$ used in Eq. (6), the result is

$$\zeta_{\text{opt.}} = \frac{\sigma}{2} \left( \Delta\alpha - \eta\mathcal{A} \left( 1 + \frac{\Delta\alpha - \Delta\alpha_+}{\mathcal{A}} \right) \right). \quad (C3)$$

Again, the relative difference is of the order $\frac{\Delta\alpha - \Delta\alpha_+}{\mathcal{A}}$.

The model from [30] still neglects the role of ions by setting $\Delta\alpha_+ = 0$. For example, the contribution of $\Delta\alpha_+$ is estimated to be more than the half of $\Delta\alpha$ for HHG in Ar driven by a 270 nm laser, which was obtained by ab initio calculations as shown in the supplement of [12]. This would indicate that the approach without using the correction for depopulation of neutrals might be even more accurate in these conditions. Moreover, to our best knowledge, the polarizability of ions has not been thoroughly tabulated and investigated in detail for a large range of wavelengths; furthermore, a precise treatment of the ultrafast phenomena may formally need a dynamical treatment [32]. However, as shown above, these differences are usually negligible in HHG-schemes, so even a simple phase-matching model considering only dispersion of neutrals and free electrons, which was used throughout this manuscript, provides sufficient accuracy.

### D. IONIZATION DEGREE OF THE MEDIUM

The choice of the initial ionization degree $\eta$ is not treated in detail in our work. The reason is that $\eta$ is a *free parameter* in our general design. In real experimental conditions, the ionization depends mainly on the driving-pulse intensity. This would need to be considered in a concrete design implementing the scheme introduced in the paper. Here we show that our optimized monochromatizing scheme works well for a broad range of $\eta$ fulfilling $\eta < \eta_c$ ($\eta_c$ being the level satisfying the perfect phase-matching condition for plane waves introduced below Eq. (6), see [31] for more details).

To illustrate the role of various values of ionization degree, we show analogies of Fig. 6 for $\eta = 0.5\eta_c$ and $\eta = 0.75\eta_c$ in Figs. 12 and 13, respectively. For those examples, we have recalculated the optimal geometrical phase $\zeta$ of the BG beam and $\xi$ (the ratio of lengths of the empty space and the segment with gas), while all the other parameters remained the same. Similarly to Fig. 6, the red crosses indicated the values obtained by the optimization algorithm.

These results show that, upon optimization, there is no significant difference in both the harmonic signal and the contrast for the different values of $\eta$ (the color bar ranges of all the figures are very similar). It is apparent that for a given configuration the trade-off between signal and contrast does not depend on the ionization degree $\eta$.

On the other side, the ionization degree clearly affects the optimized design, as we see the effect of different initial $\eta$ on the optimized parameters $\zeta \approx \frac{\theta^2}{2}$ and $\xi$. According to Eq. (A4a), the main idea of our optimized design is that the optimal phase matching within a single period is given by the sum of wave number contributions of the ionization through $\eta$ and the BG beam characterized by $\theta$. Therefore, higher values of $\eta$ lead to the decrement of $\theta$ as indicated by the red crosses in the figures. Consequently, $\xi$ is incremented to ensure the accumulation of the proper phase jump in the "empty" segment. Namely for the currently discussed case, $\xi_{\eta=0.25\eta_c} \approx 0.87$ (the value used in the main manuscript), $\xi_{\eta=0.5\eta_c} \approx 1.3$, and $\xi_{\eta=0.75\eta_c} \approx 2.6$. All these values seem realistic for a potential experimental implementation.

We note that the basic monochromatizing condition introduced in Sec. 2 works also for $\eta > \eta_c$. However, the generation within a single period of the medium cannot be phase-matched anymore. This is caused by the fact that the negative contribution to the wave number from the geometrical aspect of focusing cannot compensate for the negative contribution of the neutrals and plasma when the ionization degree overcomes the critical value. Mathematically, it would require $\zeta_{opt} < 0$ in Eq. (6), which violates the geometrical constraints of BG beams.

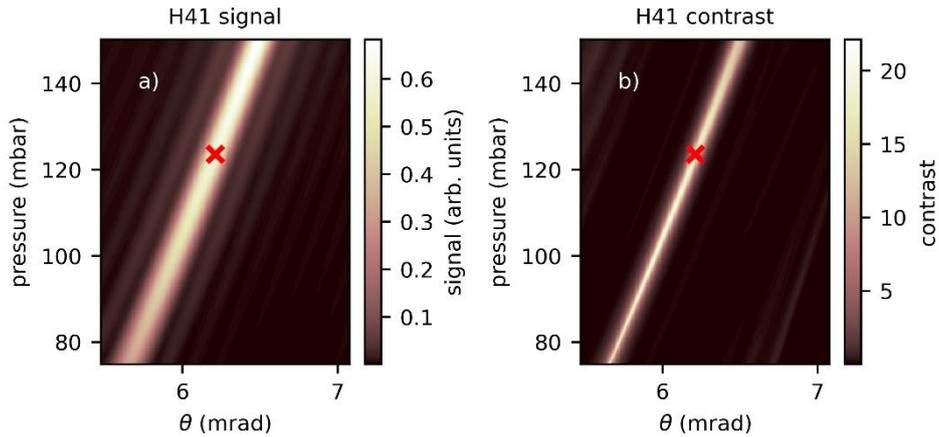

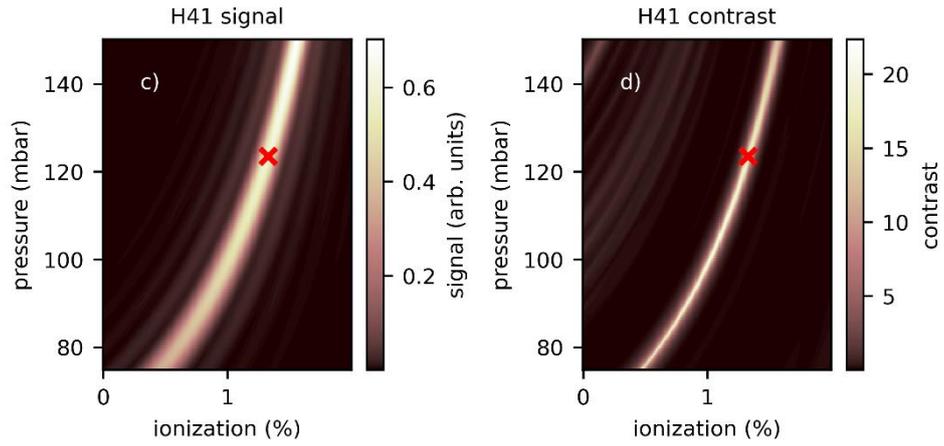

Fig. 12 - The analogy of Fig. 6 for $\eta = 0.5\eta_c$, i.e. $\eta = 1.3\%$.

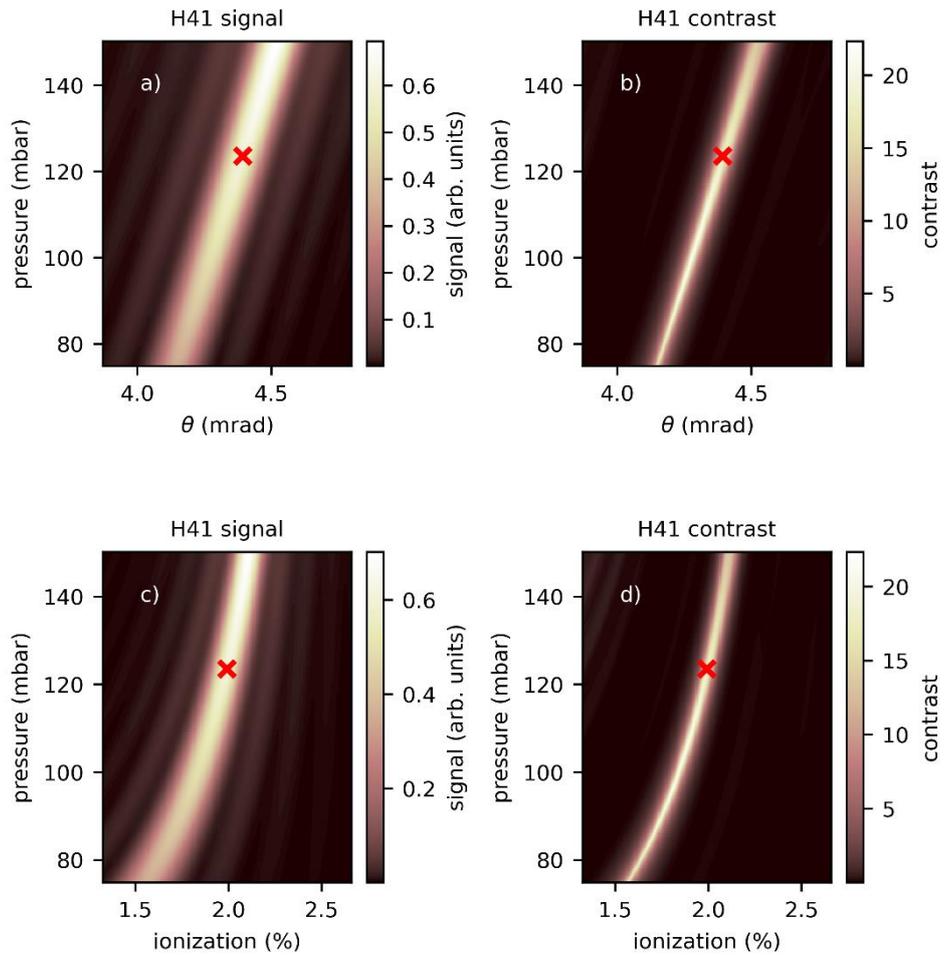

Fig. 13 - The analogy of Fig. 6 for $\eta = 0.75\eta_c$, i.e. $\eta = 2\%$.

ACKNOWLEDGMENTS

This work was supported by the project "Advanced research using high-intensity laser-produced photons and particles" (ADONIS) (CZ.02.1.01/0.0/0.0/16 019/0000789) from the European Regional Development Fund and the Ministry of Education, Youth and Sports.